\documentclass[floats,floatfix,showpacs,amssymb,prd,superscriptaddress,nofootinbib,nolongbibliography,reprint]{revtex4-2}

\usepackage{amssymb,amsmath,verbatim,mathtools,needspace,enumitem,etoolbox,graphicx,physics,microtype,afterpage,xspace,tabularx,lmodern,multirow}
\usepackage[utf8]{inputenc}
\usepackage{comment}
\usepackage[margin=2.5cm]{geometry}
\usepackage{graphicx}
\usepackage{caption}
\usepackage{subfig} 
\usepackage[dvipsnames]{xcolor}
\usepackage{multirow}
\usepackage{amsmath}
\usepackage{amssymb}
\usepackage{amsthm}
\usepackage{mathrsfs}
\usepackage[colorlinks=true,linkcolor=red]{hyperref}
\usepackage{tabularx}

\usepackage{ulem}

\newcommand\prlsec[1]{\vspace{2mm}\noindent {\textbf{  \textit{#1}}}}

\begin{document}

\title{{ 
Radial perturbations of charged wormholes}}

\author{Jose Luis Bl\'azquez-Salcedo} 
\author{Luis Manuel Gonz\'alez-Romero} 
\author{{Fech Scen Khoo}}
\affiliation{Departamento de F\'isica Te\'orica and IPARCOS, Facultad de Ciencias F\'isicas, Universidad Complutense de Madrid, Spain}
\author{{Jutta Kunz}}
\affiliation{Institut f\"ur  Physik, Universit\"at Oldenburg, Postfach 2503, D-26111 Oldenburg, Germany}
\author{{Pablo Navarro Moreno}}
\affiliation{Departamento de F\'isica Te\'orica and IPARCOS, Facultad de Ciencias F\'isicas, Universidad Complutense de Madrid, Spain}

\date{\today}

\begin{abstract}

Ellis-Bronnikov wormholes suffer from an unstable radial mode.
Here we investigate the evolution of the unstable mode(s) for charged wormholes.
We show that the instability remains in the presence of charge, but exhibits a very fast decrease to zero.
{
We hereby make a full study of the spectrum of the unstable radial modes. 
}
For so-called supercritical wormholes, two purely imaginary unstable modes merge and continue with degenerate imaginary parts and opposite real parts.
By analogy, we conjecture an analogous behavior for rotating chargeless wormholes.
\end{abstract}

\maketitle

\prlsec{Introduction.}
In general relativity (GR), traversable Lorentzian wormholes either need an unknown form of matter, so-called exotic matter \cite{Morris:1988cz,Visser:1995cc,Ellis:1979bh}, or quantum matter \cite{Blazquez-Salcedo:2020czn,Konoplya:2021hsm,Blazquez-Salcedo:2021udn} in order to violate the energy conditions.
A simple type of exotic matter is provided by a phantom scalar field.
Such a phantom field is employed to obtain the well-known Ellis-Bronnikov (EB) wormholes \cite{Ellis:1973yv,Bronnikov:1973fh}.
Alternative theories of gravity, on the other hand, may allow for the violation of the energy conditions in the gravitational sector and thus do without exotic matter or quantum matter \cite{Kanti:2011jz,Harko:2013yb,Bakopoulos:2021liw} (for a review see, e.g., \cite{Lobo:2017cay}).

The original EB wormholes are static spherically symmetric solutions.
However, the inclusion of rotation seems rather relevant from an observational point of view, since rotation is ubiquitous in the universe.
Rotating EB wormholes were first obtained perturbatively for slow rotation \cite{Kashargin:2007mm,Kashargin:2008pk} and later for rapid rotation \cite{Kleihaus:2014dla,Chew:2016epf},
where, however, closed form solutions have not yet been found \cite{Volkov:2021blw}.

An intriguing open question concerns the stability of rotating EB wormholes.
The static EB wormholes possess a radial instability
\cite{Shinkai:2002gv,Gonzalez:2008wd,Gonzalez:2008xk,Cremona:2018wkj,Xu:2025jad}.
Therefore, the question is whether the presence of rotation would stabilize these wormholes \cite{Matos:2010pcd}.
Our previous studies of the radial modes of slowly rotating EB wormholes have hinted at stabilization, since the notoriously unstable mode of the static wormholes became less unstable \cite{Azad:2023iju,Azad:2024axu}.

However, at the same time a second radial mode evolves from the static EB wormhole, that becomes more unstable with increasing rotation.
In perturbation theory both unstable modes simply cross.
But as shown before for rotating EB wormholes in five dimensions, the modes actually merge at a critical value of the angular momentum.
Beyond, these purely imaginary modes were no longer seen \cite{Dzhunushaliev:2013jja}.

To clarify the question of radial in/stability, a numerical analysis of the radial modes of rapidly rotating wormholes would be necessary, employing spectral methods \cite{Blazquez-Salcedo:2023hwg,Khoo:2024yeh,Blazquez-Salcedo:2024oek, 
Blazquez-Salcedo:2024dur,Khoo:2024agm}.
Since this still represents a challenge, we here resort to a simpler problem, considering charge instead of angular momentum, and conclude by analogy.
Stability and properties of other types of (charged) wormholes have also been studied e.g. in \cite{Gonzalez:2009hn,Bronnikov:2012ch,Torii:2013xba,Nozawa:2020wet,DeFalco:2023twb,Ilyas:2023rde,Jaramillo:2023pny,Bronnikov:2025zgw}.

Here we show that for a set of charged EB wormholes, that are known in closed form \cite{Gonzalez:2009hn}, analogous to the rotating case, two branches of purely imaginary modes bifurcate at a critical solution.
However, the unstable modes do not disappear then.
Instead, at the bifurcation the modes develop a real part with increasing charge, yielding two modes with equal imaginary parts but opposite real parts.
As the charged EB wormholes approach the limiting solution, i.e., the exterior metric approaches the metric of an extremal Reissner-Nordstr\"om black hole, the degenerate imaginary part of the modes tends very fast
to zero.

\prlsec{Theoretical framework.}
We consider the Einstein-Maxwell-scalar action
\begin{equation}
    \mathcal{S}=\frac{1}{16\pi}\int d^4x\sqrt{-g}\left[ \mathrm{R}-F^2+2(\nabla\varphi)^2\right],
\end{equation}
where $\mathrm{R}$ is the Ricci scalar, $F$ is the electromagnetic tensor, and $\varphi$ is a phantom scalar field. 
When varying the action with respect to the metric $g$, the gauge field $A$, and the scalar field $\varphi$, the equations of motion follow.

The background solution of the charged EB wormholes was obtained by Gonz\'alez et al.~\cite{Gonzalez:2009hn}.
It may be expressed in the form
\begin{eqnarray}
     ds^2 &=&  -G(r) dt^2 
     +{\frac{1}{G(r)}} \times 
     \nonumber \\ &&
    \left[dr^2+(r^2+r_0^2) 
 {   (d\theta^2+ \sin^2{\theta}  \,
    d\phi^2 }
    ) \right] \,,
    \label{eqm}
\end{eqnarray}
\begin{equation}
    \varphi=\varphi_1\arctan{(r/r_0)}+\varphi_0 \, ,
    \label{eqa}
\end{equation}
\begin{equation}
    F=\frac{2Q_e}{r^2+r_0^2}G(r)  dt\wedge dr+2Q_m\sin{\theta}d\theta\wedge d\phi \, ,
    \label{eqp}
\end{equation}
with electromagnetic charges $Q_e$ and $Q_m$, 
and a function $G$.

Gonz\'alez et al.
distinguish three cases for the charged wormhole solutions:
the subcritical case with $\Lambda>0$, the critical case with $\Lambda=0$, and the supercritical case with $\Lambda=i\mu$ with a real positive number $\mu$.
The function $G(r)$ changes accordingly for these cases with a parameter $\gamma_1$
\cite{Gonzalez:2009hn},
\begin{equation}
G(r)
= \left\{
\begin{array}{cc} 
 {\sigma^2} \left[\cosh{(\Lambda y)}-\gamma_1\frac{\sinh{(\Lambda y)}} {\Lambda}\right]^{{-2}}  \, ,
& \Lambda > 0 \, ,\\[5pt]
 {\sigma^2} (1-\gamma_1y)^{{-2}} \, ,
& \Lambda = 0 \, ,\\[2pt]
{\sigma^2} \left[ \cos(\mu y) - \gamma_1\frac{\sin(\mu y)}{\mu} \right]^{{-2}} \, ,
& {\mu >0} \, ,
\end{array} 
\right.
\label{cases1}
\end{equation}
where $y=\arctan{(r/r_0)}$, {and the parameter $\sigma$ is chosen {such} that $g_{tt}=-1$, $g_{rr}=1$ when $r {\rightarrow} \infty$}. 
In order to obtain global wormhole solutions, the conditions 
\begin{equation}
\left.
\begin{array}{cc}
  {\Lambda < |\gamma_1|} \hbox{ and }
  \frac{\tanh\left(\Lambda\frac{\pi}{2}\right)}{\Lambda} |\gamma_1| < 1 \, ,
& \hbox{subcritical case},\\[5pt]
  \frac{\pi}{2}|\gamma_1| < 1 \, ,
& \hbox{critical case},\\[2pt]
  \mu < 1 \hbox{ and }
  \frac{\tan\left(\mu\frac{\pi}{2}\right)}{\mu} |\gamma_1| < 1 \, ,
& \hbox{supercritical case} .
\end{array} 
\right.
\label{cases2}
\end{equation}
should hold.
The chargeless EB solution 
is found when $\Lambda=\gamma_1$.

We restrict our calculations to the purely electric case, i.e., $Q_{m}=0$.
Setting $\varphi_1=Q_s/r_0$, the asymptotic expansion of $\varphi$ identifies the scalar charge $Q_s$, and $\varphi_0=-\pi Q_s/2 r_0$ fixes the scalar asymptotic value. 
Furthermore, the field equations yield the following relations between the parameters, 
\begin{equation}
Q_s = \left\{
\begin{array}{cc}
r_0\sqrt{1+\Lambda^2} \, , & \rm{subcritical},\\[1ex]
r_0 \, , & \rm{critical} ,\\[1ex]
r_0\sqrt{1-\mu^2} \, , & \rm{supercritical},
\end{array}
\right.
\label{cases3}
\end{equation}
and finally
$\gamma_1=\big(\sqrt{Q_s^2+{\sigma}^{2} Q_e^2-r_0^2}\big)/r_0$.
Hence the wormholes are characterized only by three parameters: the charges $Q_e$, $Q_s$ and the scale $r_0$. 

\begin{figure}[t!]
\begin{center}
\resizebox{1\columnwidth}{!}{
  \includegraphics[angle=-90]{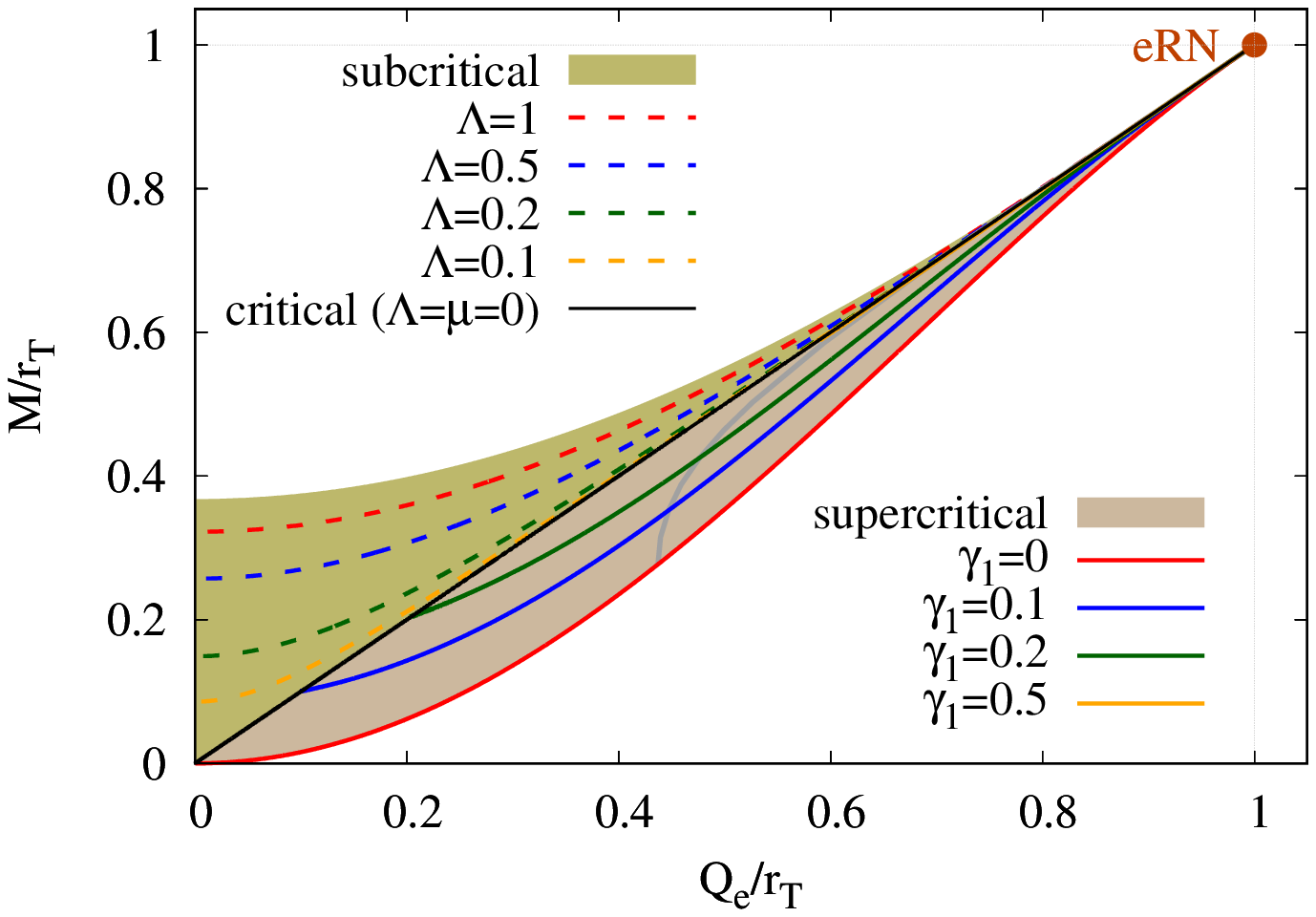}
}
\vspace*{-0.7cm}
\end{center}
\caption{
Domain of existence 
of the electrically charged wormholes: The brown shaded area corresponds to supercritical wormholes, 
while the green shaded area corresponds to the subcritical wormholes. 
They are separated by the critical wormholes 
(black line). Other 
colored 
curves correspond to families of solutions with fixed values of the parameters. 
All the curves terminate at the extremal Reissner-Nordström black hole (eRN).
}
\label{fig1}
\end{figure}

However, besides $Q_e$ the physically meaningful quantities are the mass of the wormholes $M$, and the circumferential radius of the throat, $r_T=\sqrt{A_T/4\pi}$ where $A_T$ is the area of the wormhole throat given by the metric (\ref{eqm}). Therefore,
we exhibit the domain of existence of supercritical,
critical and subcritical 
wormholes in Figure \ref{fig1}, 
showing
the mass $M$ and the electric charge $Q_e$ scaled with the radius of the throat $r_T$.
Critical solutions satisfy $Q_e=M$, supercritical solutions have $Q_e>M$, and for subcritical $Q_e<M$.
Analysis of the wormhole mass
at both infinities reveals that $\gamma_1$ plays the role of the asymmetry parameter \cite{Gonzalez:2009hn}. For $\gamma_1=0$, symmetric wormhole solutions result,
and they form the lower boundary of the supercritical domain (red curve).
Finally, note that the electric charge and the mass are bounded by $Q_e=M=r_T$, precisely where the extremal Reissner-Nordström (eRN) black hole is located ($r_T\to r_H$, 
$r_H$ being the location of the black hole horizon).

\prlsec{Radial perturbations.}
We now turn to the radial perturbations.
In general, the perturbed fields read
\begin{equation}
    g_{\mu\nu}=g_{\mu\nu}^0+\delta g_{\mu\nu} \, ,
    \ \ 
    A_{\mu}=A_{\mu}^0+\delta A_{\mu} \,,
    \ \ 
    \varphi=\varphi^0+\delta\varphi \,,
\end{equation}
where the superscript $0$ denotes the background solution given by Eqs.~\eqref{eqm} - \eqref{eqp}.
The perturbations
read for the metric
\begin{eqnarray}
 && \delta g_{\mu\nu}(t,r,\theta,\phi)= e^{-i\omega t} \times
 \Bigl(
 G(r)F_0(r) dt^2 + 
 \nonumber \\
 &&  
 \frac{F_1(r)}{G(r)} dr^2 + \frac{F_2(r)}{G(r)} (r^2+r_0^2)d\Omega^2  
 \Bigl) \, ,
\end{eqnarray}
and for the vector and scalar fields,
\begin{eqnarray}\label{deltaA_polar}
\delta A(t,r,\theta,\phi)&=& e^{-i\omega t} \left( V_0(r)dt  
+ V_1(r)dr   
\right) \, ,
\ \ \ \\
\delta\varphi(t,r,\theta,\phi)&=&u(r)e^{-i\omega t} \, ,
\end{eqnarray}
where $\omega$ is an eigenfrequency.
Note that $\omega=\omega_R+i\omega_I$ is a complex number in general. Unstable modes are characterized by $\omega_I>0$, and we can define the 
{
characteristic time of the instability
}
as $\tau_0=1/\omega_I$.

To fix the gauge, we set $F_0+F_1-2F_2=0$  
and $V_1 = F_0 \, \partial_r A_t^0/\omega$.
This allows us to decouple $F_1$ and $V_0$ 
in
the resulting system of differential equations, 
hence
decreasing its order.
The final set of equations to solve is
\begin{eqnarray}\label{eq_F0_radial}
 &&  H F_0' =  \left[ 2G^2Q_e^2 - 2Gr_0^2 - rR^2G'\right] F_0 \\
 && \:   +    2\left[ G^2 r_0^2 - 
 R^4 G^{-1}
 \omega^2 \right] F_2  -4G Q_s \left[ ru + R^2 u' \right] \, ,  \nonumber
\end{eqnarray}
\begin{equation}\label{eq_F2_radial}
   F_2'=
    R^{-2}(2 Q_s u + 2 r) F_2
   + R^{-4}G^{-1} H F_0 /2 \, ,
\end{equation}
\begin{equation}\label{eq_u_radial}
    u''=-2rR^{-2}u'-G^{-2}\omega^2 u \, ,    
\end{equation}
where $R^2=r^2+r_0^2$ and $H=-R^2G^2(R^2/G)'$.
We are interested in the solutions 
to
these equations that behave asymptotically as outgoing waves when we move far away from the throat, 
i.e.
\begin{equation}
 {   F_0(r) \sim re^{i\omega r^*}\,, F_2(r) \sim e^{i\omega r^*}\,, u(r) \sim 
 r^{-1}
 e^{i\omega r^*}\,,  }
\end{equation}
when $r\to\pm\infty$,
where the tortoise coordinate is
\begin{equation}\label{tortoise}
    \frac{dr^*}{dr}=
G^{-1}\, .
\end{equation}
Finally let us mention that the previous equations can be cast into a single master equation (Schrödinger-like). The problem is that the resulting potential happens to be divergent at the throat, and some additional transformation has to be employed in order to regularize the master equation (see e.g. \cite{Gonzalez:2009hn, Bronnikov:2012ch, Torii:2013xba}). However, we will avoid this problem by directly integrating the previous system of differential equations (\ref{eq_F0_radial})-(\ref{eq_u_radial}), which is regular at the throat.

\prlsec{Spectral method.}

In order to calculate the spectrum of quasinormal modes for the radial perturbations, we employ a spectral method. 
We introduce a compactified radial coordinate, $x= (2/\pi) y$ for which $x(r=\pm\infty)=\pm1$.
Then we decompose the perturbations 
$F_i=\{F_0,F_2,u\}$
into 
\begin{equation}\label{spectral_decomp}
    F_i(x)=\sum_{k=0}^{N_p-1}C_{i,k}T_k(x) \, ,
\end{equation}
where $N_p$ is the size of the grid, 
$C_{i,k}$ are the coefficients of the decomposition for the $i$-th function and $T_k(x)$ denotes the Chebyshev polynomials of the first kind.
We discretize the domain by choosing a Gauss-Lobatto grid in $x$, casting the system as a quadratic eigenvalue problem:
\begin{equation}
    \left(\mathcal{M}_0+\mathcal{M}_1\omega+\mathcal{M}_2\omega^2\right)C=0 \, ,
\end{equation}
with $(3\times N_p) \times (3\times N_p)$ matrices $\mathcal{M}_0$, $\mathcal{M}_1$, and $\mathcal{M}_2$
and a vector of constants $C$, containing the coefficients $C_{i,k}$ of the decomposition.
From here we extract the quasinormal modes, making use of
Matlab with the Advanpix Multiprecision Computing Toolbox \cite{Advanpix} (see \cite{Blazquez-Salcedo:2023hwg,Khoo:2024yeh,Blazquez-Salcedo:2024oek, 
Blazquez-Salcedo:2024dur,Khoo:2024agm} for more details). 
The modes are obtained with a very high precision, i.e., typically with estimated errors of the order of  $\sim 10^{-6}$ or better.

\begin{figure}[t!]
\begin{center}
\resizebox{1\columnwidth}{!}{
    \includegraphics[angle=-90]{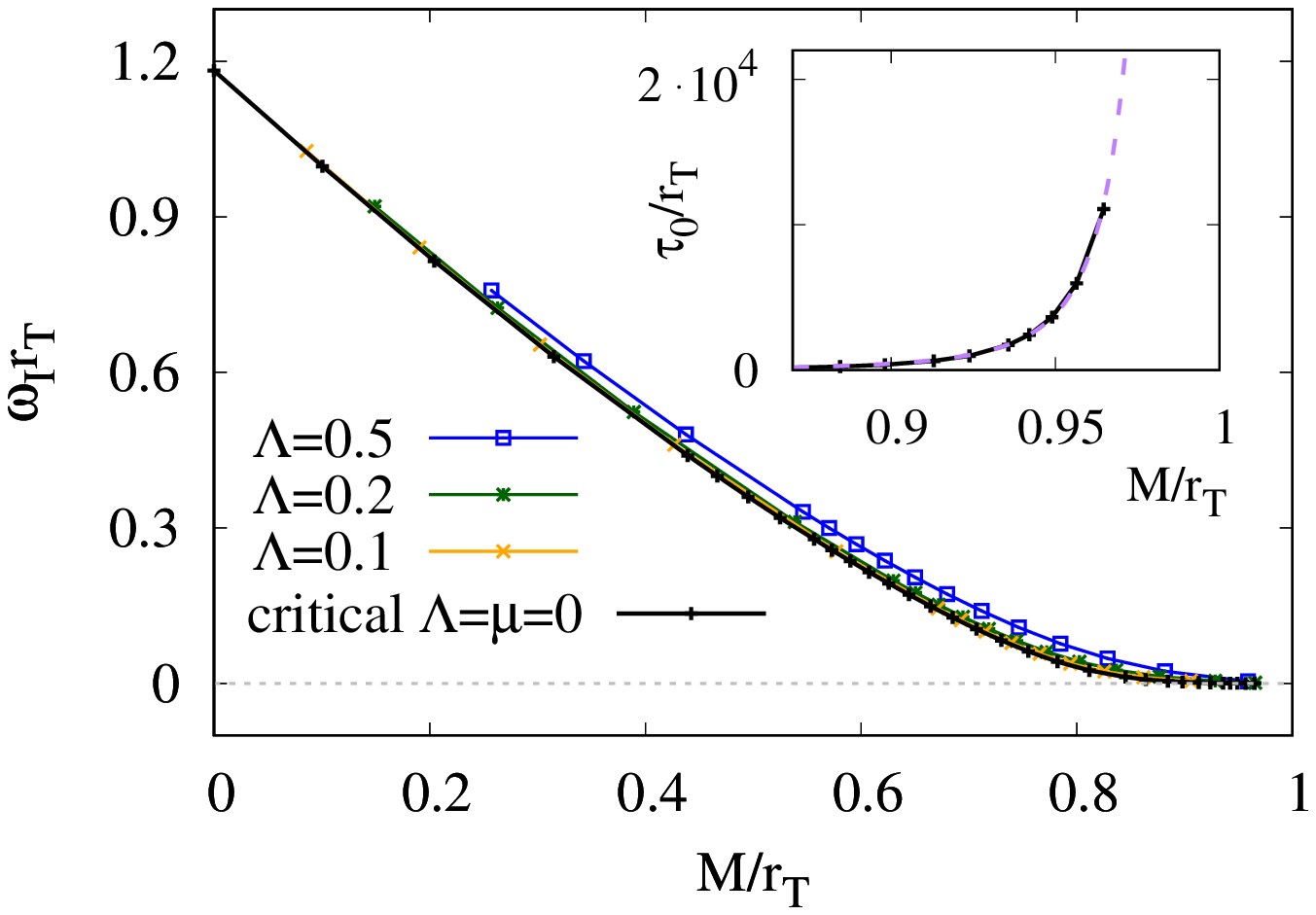}
}
\vspace*{-0.7cm}
\end{center}
\caption{Unstable radial modes of subcritical and critical wormholes. The inset shows 
the critical wormhole decay time $\tau_0$ as a function of the mass $M$,
and a fit (dashed line) 
described by 
(\ref{num_rel}).
}
\label{fig2}
\end{figure}

\prlsec{Results.}
We now present the results of the numerical calculations of the radial modes, starting with the subcritical case. The analysis of the spectrum reveals that there is an unstable mode for this type of solutions. The mode is purely imaginary with $\omega=i\omega_I$. In Figure \ref{fig2} we show the mode as a function of the mass, scaled with the throat radius. The curves in color correspond to families of solutions with a fixed value of $\Lambda$.
For comparison,
in black we show the family of critical solutions ($\Lambda=0$), for which the mass exactly equals  the electric charge. The subcritical 
modes are in fact very close to the critical 
modes,
showing that the dependence of the {subcritical} modes on $\Lambda$ is small.
The value of both the critical and subcritical modes decreases monotonically as we increase 
the scaled mass. 
Furthermore, the modes become extremely small as we approach the limit of the solutions, i.e., the eRN black hole with $r_T=r_H=M=Q_e$. %

An analysis of the numerical data shows that 
the decay of the critical modes (in black) follows a relation,
\begin{equation}
    \omega_I\propto (1-M/r_T)^{-3.1} \, ,
    \label{num_rel}
\end{equation}
satisfied within a $1\%$ error.
This means that by moving closer to the eRN solution, we can increase the 
{
characteristic time of the instability $\tau_0$
}
to arbitrarily large values.

\begin{figure}[t!]
\begin{center}
\resizebox{1\columnwidth}{!}{
    \includegraphics[angle=-90]{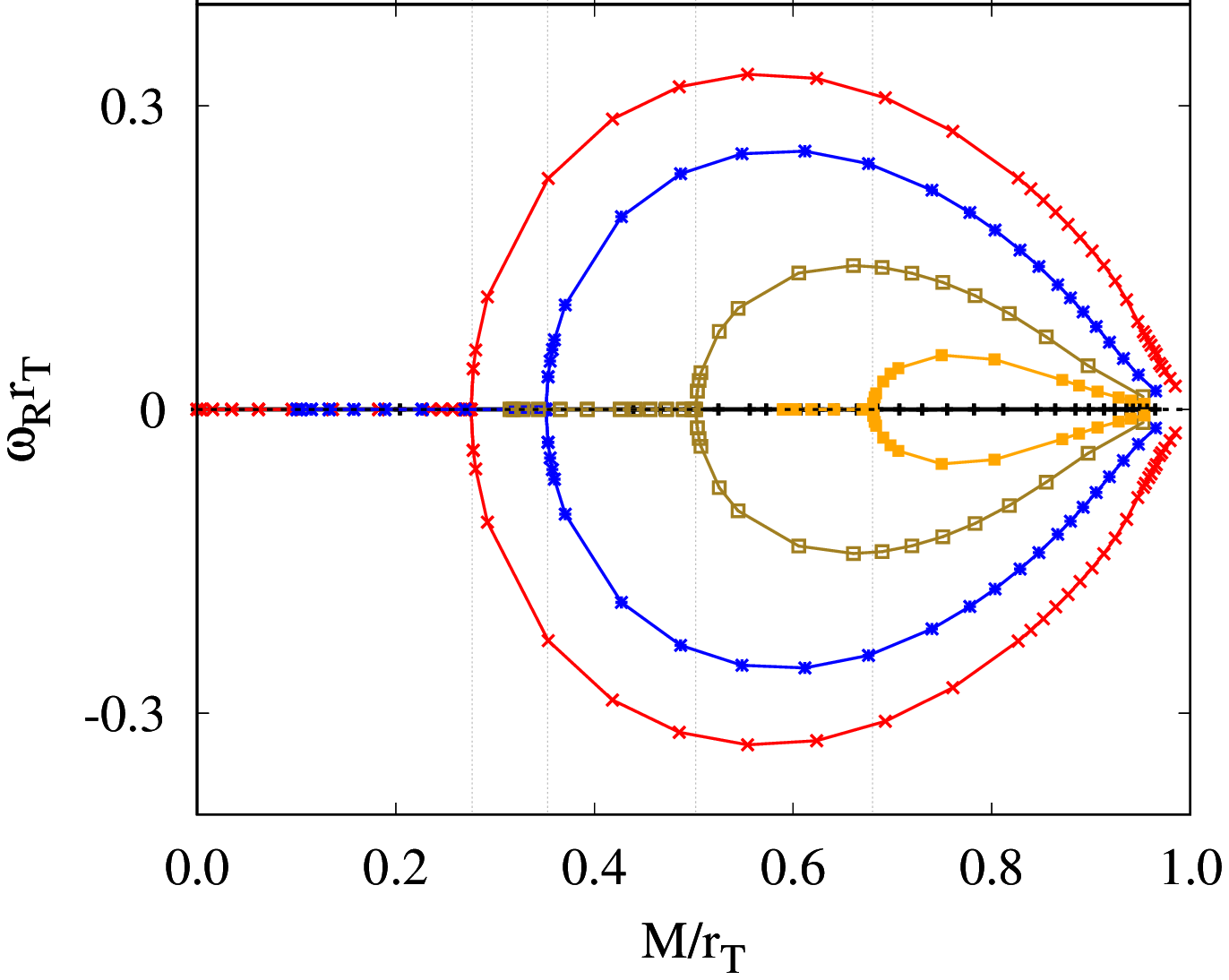}
}
\vspace*{-0.7cm}
\end{center}
\caption{Unstable radial modes of supercritical and critical wormholes.
}
\label{fig3}
\end{figure}

Next we focus on the supercritical wormhole solutions.
In the upper panel of Figure \ref{fig3} 
we exhibit the imaginary part $\omega_I$ of the radial modes, scaled with the circumferential throat radius $r_T$, versus the scaled mass $M/r_T$ of the wormholes for several choices of the parameter $\gamma_1$. 
A similar figure is shown in the lower panel for the real part $\omega_R$ of the modes.
The radial instability of this set of solutions appears to be more involved, as  the mode dependence evolves differently, compared to the critical solution shown here as well in black.

We observe for each $\gamma_1$ the same pattern:
for small masses two purely imaginary modes exist, i.e., they have $\omega_R=0$.
The more unstable mode 
(with a larger $\omega_I r_T$ value) 
then gets less unstable with increasing $M/r_\text{T}$, while the unstable mode emerging from a zero mode becomes more unstable.
Both modes then merge at a critical value of the mass and charge.
In Figure \ref{fig1} we mark these critical values with a solid grey line within the supercritical region.

However, whereas one might be tempted to conjecture that the radial instability disappears at the critical value, this is not the case.
Instead, a bifurcation arises at the critical value, where the two modes obtain non-vanishing real parts $\pm \omega_R$.
This bifurcation is a novel result that has not been observed in \cite{Gonzalez:2009hn}.
The instability continues beyond the bifurcation with two modes that possess the same imaginary part but opposite real parts.
As we increase the value of $M/r_\text{T} \to 1$, we reach 
the eRN solution.
The figure clearly shows that in this limit the radial instability finally disappears.
Intriguingly, the imaginary part $\omega_I$ decreases to zero very fast, following a relation similar to the one obtained in the critical case (see (\ref{num_rel})).
Thus, 
like previously for the subcritical case,
while not yet fully radially stable, it is 
possible to obtain an arbitrarily long 
{
characteristic time
}
$\tau_0$ 
with an increasing
$M/r_T$ and $Q/r_T$ close to the eRN value. Hence this indicates that in the region of solutions close to the eRN limit, 
{
the linear radial perturbations of the wormholes can take a very long time to grow.
Of course this does not mean that other instabilities 
would not creep up, for example in the non-linear regime.}

To contrast
the order of magnitude of 
{
the characteristic time of the instability,
}
let us consider the case of a critical wormhole with $r_T=10$ km. In the uncharged case, this is simply the massless EB solution with an extremely short
{
characteristic time,
}
$\tau_0 = 28$  $\mu$s. 
As we noted before, a highly 
charged wormhole can 
{
possess an arbitrarily long $\tau_0$.
}
For example, for $Q_e/r_T=0.95$ the critical wormhole would 
{
have
}
$\tau_0=0.13$ s
and for $Q_e/r_T=0.9999$ the {
characteristic time of the instability
}
would be around $1.1$ years.

Finally let us draw an analogy to the rotating chargeless EB wormholes.
Although the instability analysis has only been done in the slow rotation regime \cite{Azad:2023iju,Azad:2024axu}, the slowly rotating wormholes have two radially unstable modes, similar to the supercritical charged wormholes. The extrapolation of the slow rotation result showed that the two modes 
merged
around $1/3$ of the maximum value of the scaled angular momentum, 
at
which the extremal Kerr black hole is found. 
This behavior of the modes 
with respect to
the angular momentum 
resembles
the behavior of the modes presented here with respect to the electric charge in the case of supercritical wormholes (see for example, the symmetric $\gamma_1=0$ wormholes in Figure \ref{fig3}). 
It is strongly suggestive for an analogous tendency to occur
close to extremality
for sufficiently fast rotating chargeless
EB
wormholes (beyond the perturbative analysis).
Hence, the  
{
characteristic time of the instability 
}
of the rotating wormholes can possibly be long, close to the extremal Kerr solution.

\prlsec{Conclusions.}
EB wormholes are well-known for their notorious radial instability.
Here we have addressed the unstable radial modes of charged EB wormholes, that come in three variants, subcritical, critical and supercritical.
The unstable radial modes of all variants exhibit a fast decay to zero, as the wormhole metric tends to the metric of an eRN black hole.
{
As a result, the characteristic time of the instability can be arbitrarily long.
}
Furthermore this also brings us to an interesting quest on the (effective) 
{
radial
}
stability of other charged wormhole solutions such as those in Einstein-Maxwell-Dirac theory \cite{Blazquez-Salcedo:2020czn,Konoplya:2021hsm,Blazquez-Salcedo:2021udn}. 
They possess $Q_e>M$, like the supercritical charged EB wormholes, and approach the eRN black hole for large enough values of the electric charge as well.

Besides the wormhole charge, another natural parameter is the angular momentum where in this case the rotating chargeless EB wormhole metric tends to the extremal Kerr metric.
An analogy can be drawn most closely to the
supercritical case for the static charged wormholes, where two unstable purely imaginary modes bifurcate and continue with degenerate imaginary parts and opposite real parts.
For the rotating chargeless EB wormholes,
a perturbative treatment in the angular momentum has also seen two purely imaginary modes arise and meet \cite{Azad:2023iju,Azad:2024axu}. 
The fate of the unstable modes beyond the meeting point, whether they will take an analogous route like the supercritical wormholes can only be confirmed by a non-perturbative study in the future.


\prlsec{Acknowledgment.}
We gratefully acknowledge support by MICINN project PID2021-125617NB-I00 ``QuasiMode''.
JLBS gratefully acknowledges support from MICINN project CNS2023-144089 ``Quasinormal modes''.
FSK gratefully acknowledges support from ``Atracci\'on de Talento Investigador Cesar Nombela'' of the Comunidad de Madrid under the grant number 2024-T1/COM-31385.
PNM gratefully acknowledges support from Universidad Complutense de Madrid through ``Contratos predoctorales
de personal investigador en formación CT25/24'' and IPARCOS under ``Ayudas de doctorado IPARCOS-UCM/2024''.


\begin{thebibliography}{99}

\bibitem{Morris:1988cz}
M.~S.~Morris and K.~S.~Thorne,
Am. J. Phys. \textbf{56}, 395 (1988)


\bibitem{Visser:1995cc}
M.~Visser,
``Lorentzian wormholes: From Einstein to Hawking,''
Woodbury, USA: AIP (1995) 


\bibitem{Ellis:1979bh}
H.~G.~Ellis,
Gen. Rel. Grav. \textbf{10}, 105 (1979)




\bibitem{Blazquez-Salcedo:2020czn}
J.~L.~Bl\'azquez-Salcedo, C.~Knoll and E.~Radu,
Phys. Rev. Lett. \textbf{126}, 101102 (2021)

\bibitem{Konoplya:2021hsm}
R.~A.~Konoplya and A.~Zhidenko,
Phys. Rev. Lett. \textbf{128}, 091104 (2022)

\bibitem{Blazquez-Salcedo:2021udn}
J.~L.~Bl{\'a}zquez-Salcedo, C.~Knoll and E.~Radu,
Eur. Phys. J. C \textbf{82}, 533 (2022)




\bibitem{Ellis:1973yv} 
  H.~G.~Ellis,
  J.\ Math.\ Phys.\  {\bf 14}, 104 (1973)

\bibitem{Bronnikov:1973fh} 
  K.~A.~Bronnikov,
  Acta Phys.\ Polon.\ B {\bf 4}, 251 (1973).

\bibitem{Kanti:2011jz}
  P.~Kanti, B.~Kleihaus and J.~Kunz,
  Phys.\ Rev.\ Lett.\  {\bf 107}, 271101 (2011)


\bibitem{Harko:2013yb}
T.~Harko, F.~S.~N.~Lobo, M.~K.~Mak and S.~V.~Sushkov,
Phys. Rev. D \textbf{87}, 067504 (2013)

\bibitem{Bakopoulos:2021liw}
A.~Bakopoulos, C.~Charmousis and P.~Kanti,
JCAP \textbf{05}, 022 (2022)


\bibitem{Lobo:2017cay}
F.~S.~N.~Lobo,
Fundam. Theor. Phys. \textbf{189}, pp.-279 (2017)
Springer, 2017

\bibitem{Kashargin:2007mm}
P.~E.~Kashargin and S.~V.~Sushkov,
Grav. Cosmol. \textbf{14}, 80 (2008)

\bibitem{Kashargin:2008pk}
P.~E.~Kashargin and S.~V.~Sushkov,
Phys. Rev. D \textbf{78}, 064071 (2008)

\bibitem{Kleihaus:2014dla}
B.~Kleihaus and J.~Kunz,
Phys. Rev. D \textbf{90}, 121503 (2014)

\bibitem{Chew:2016epf}
X.~Y.~Chew, B.~Kleihaus and J.~Kunz,
Phys. Rev. D \textbf{94}, 104031 (2016)

\bibitem{Volkov:2021blw}
M.~S.~Volkov,
Phys. Rev. D \textbf{104}, 124064 (2021)

\bibitem{Shinkai:2002gv}
  H.~a.~Shinkai and S.~A.~Hayward,
  Phys.\ Rev.\ D {\bf 66}, 044005 (2002)

\bibitem{Gonzalez:2008wd}
  J.~A.~Gonz\'alez, F.~S.~Guzman and O.~Sarbach,
  Class.\ Quant.\ Grav.\  {\bf 26}, 015010 (2009)

\bibitem{Gonzalez:2008xk}
  J.~A.~Gonz\'alez, F.~S.~Guzman and O.~Sarbach,
  Class.\ Quant.\ Grav.\  {\bf 26},  015011 (2009)

\bibitem{Cremona:2018wkj}
F.~Cremona, F.~Pirotta and L.~Pizzocchero,
Gen. Rel. Grav. \textbf{51}, 19 (2019)

\bibitem{Xu:2025jad}
A.~Xu, X.~Y.~Chew and D.~h.~Yeom,
JCAP \textbf{08}, 012 (2025)
 
\bibitem{Matos:2010pcd}
T.~Matos,
Gen. Rel. Grav. \textbf{42}, 1969 (2010)

\bibitem{Azad:2023iju}
B.~Azad, J.~L.~Bl\'azquez-Salcedo, F.~S.~Khoo and J.~Kunz,
Phys. Lett. B \textbf{848}, 138349 (2024)

\bibitem{Azad:2024axu}
B.~Azad, J.~L.~Bl{\'a}zquez-Salcedo, F.~S.~Khoo and J.~Kunz,
Phys. Rev. D \textbf{109}, 124051 (2024)

\bibitem{Dzhunushaliev:2013jja}
V.~Dzhunushaliev, V.~Folomeev, B.~Kleihaus, J.~Kunz and E.~Radu,
Phys. Rev. D \textbf{88}, 124028 (2013)

\bibitem{Blazquez-Salcedo:2023hwg}
J.~L.~Bl\'azquez-Salcedo, F.~S.~Khoo, J.~Kunz and L.~M.~Gonz\'alez-Romero,
Phys. Rev. D \textbf{109}, 064028 (2024)

\bibitem{Khoo:2024yeh}
F.~S.~Khoo, B.~Azad, J.~L.~Bl\'azquez-Salcedo, L.~M.~Gonz\'alez-Romero, B.~Kleihaus, J.~Kunz and F.~Navarro-L\'erida,
Phys. Rev. D \textbf{109}, 084013 (2024)

\bibitem{Blazquez-Salcedo:2024oek}
J.~L.~Bl{\'a}zquez-Salcedo, F.~S.~Khoo, B.~Kleihaus and J.~Kunz,
Phys. Rev. D \textbf{111}, L021505 (2025)

\bibitem{Blazquez-Salcedo:2024dur}
J.~L.~Blazquez-Salcedo, F.~S.~Khoo, B.~Kleihaus and J.~Kunz,
Phys. Rev. D \textbf{111}, 064015 (2025)


{
\bibitem{Khoo:2024agm}
F.~S.~Khoo, J.~L.~Bl{\'a}zquez-Salcedo, B.~Kleihaus and J.~Kunz,
Eur. Phys. J. C \textbf{85}, no.11, 1366 (2025)
}


\bibitem{Gonzalez:2009hn}
J.~A.~Gonzalez, F.~S.~Guzman and O.~Sarbach,
Phys. Rev. D \textbf{80}, 024023 (2009)




{
\bibitem{Bronnikov:2012ch}
K.~A.~Bronnikov, R.~A.~Konoplya and A.~Zhidenko,
Phys. Rev. D \textbf{86}, 024028 (2012)
}


\bibitem{Torii:2013xba}
T.~Torii and H.~a.~Shinkai,
Phys. Rev. D \textbf{88}, 064027 (2013)


\bibitem{Nozawa:2020wet}
M.~Nozawa,
Phys. Rev. D \textbf{103}, 024004 (2021)

\bibitem{DeFalco:2023twb}
V.~De Falco and S.~Capozziello,
Phys. Rev. D \textbf{108}, 104030 (2023)

\bibitem{Ilyas:2023rde}
M.~Ilyas and K.~Bamba,
JCAP \textbf{10}, 038 (2023)

\bibitem{Jaramillo:2023pny}
V.~Jaramillo, M.~Lira, D.~Mart{\'\i}nez-Carbajal and D.~N{\'u}{\~n}ez,
Phys. Rev. D \textbf{109}, 064007 (2024)

\bibitem{Bronnikov:2025zgw}
K.~A.~Bronnikov, S.~V.~Bolokhov, M.~V.~Skvortsova, R.~Ibadov and F.~Y.~Shaymanova,
Eur. Phys. J. C \textbf{85}, 1063 (2025)



\bibitem{Advanpix}
P. Holoborodko, 
Advanpix 5.1.0.15432,
http://www.advanpix.com


\end{thebibliography}
\end{document}